\documentclass[twocol]{ametsocV6.1}

\usepackage{gensymb}
\usepackage{siunitx}
\usepackage{multirow}
\usepackage{booktabs}
\usepackage{hyperref}
\hypersetup{
    colorlinks=true,
    linkcolor=orange,
    filecolor=magenta,
    urlcolor=cyan,
    citecolor=blue,
    pdftitle={MODIS Thermal Infrared Sounding (MOTIS): Estimating Tropical Cyclone Central Pressure from Warm-Core Anomalies},
    pdfpagemode=FullScreen,
}

\title{MODIS Thermal Infrared Sounding (MOTIS): Estimating Tropical Cyclone Central Pressure from Warm-Core Anomalies}

\authors{Jinghuai Yao\textsuperscript{*},\aff{a}\thanks{\textsuperscript{*}Jinghuai Yao, Chi Yan Kwok, and Puyuan Du contributed equally to this work. Jinghuai Yao can be reached at jyao224@wisc.edu for questions about the MOTIS dataset.}
Chi Yan Kwok\textsuperscript{*},\aff{b}
Puyuan Du\textsuperscript{*},\aff{c}
Yubo Wang,\aff{d, e}
Derrick Herndon,\aff{e}\correspondingauthor{Derrick Herndon, dherndon@ssec.wisc.edu}}

\affiliation{\aff{a}{Department of Astronomy, University of Wisconsin-Madison, Madison, Wisconsin}
\aff{b}{School of Geography, Earth and Environmental Sciences, University of Birmingham, Birmingham, United Kingdom}
\aff{c}{Department of Chemistry and Biochemistry, University of California, Los Angeles, California}
\aff{d}{Department of Atmospheric and Oceanic Sciences, University of Wisconsin-Madison, Madison, Wisconsin}
\aff{e}{Cooperative Institute for Meteorological Satellite Studies, University of Wisconsin-Madison, Madison, Wisconsin}}

\abstract{This study presents a novel framework for estimating the central sea-level pressure ($P_\mathrm{c}$) of tropical cyclones (TCs) using infrared radiometers. We leverage the long-overlooked combination of high spatial resolution and sounding capability of the Moderate Resolution Imaging Spectroradiometer (MODIS) to measure warm-core anomalies in TC eyes. We develop the MODIS Thermal Infrared Sounding (MOTIS) framework, which performs instrument-specific preprocessing and estimates $P_\mathrm{c}$ using multiple linear regression. MOTIS yields $r^2=0.945$ and $\mathrm{RMSE}=$ \qty{4.3}{\hecto\pascal} for high-intensity TCs with observed clear eyes (mean $P_\mathrm{c}=$ \qty{937}{\hecto\pascal}), outperforming all existing methods for intense TCs. We construct a dataset of 3288 (1082 clear-eye) MOTIS estimates from 2002 to 2025 and demonstrate its potential to improve the quality of Best Track $P_\mathrm{c}$, roughly halving uncertainties in the absence of pressure observations. Although MODIS is nearing the end of its mission, the MOTIS framework could be extended to next-generation geostationary sounders to provide accurate real-time $P_\mathrm{c}$ estimation for high-intensity TCs.}

\begin{document}

\maketitle

\statement
 Despite recent improvements in tropical cyclone intensity estimation, significant challenges remain in measuring the strength of the strongest storms. This study presents a new framework for estimating central sea-level pressure, a standard measure of storm intensity, from infrared satellite observations. It uses high-resolution satellite observations to measure the warm core in the eye of tropical cyclones and translate it into pressure estimates. For intense systems with clear eyes, the method roughly halves the uncertainties in historical pressure estimates. The new pressure estimate dataset from 2002 to 2025 can greatly improve storm records used for forecasting, risk assessment, and climate studies, and the approach could be extended to future satellite sounders for real-time monitoring.

\capsule
We use high-resolution MODIS infrared sounding to capture tropical cyclone warm-core anomalies and improve central pressure estimates for intense storms with clear eyes.

\section{Introduction}

The central sea-level pressure ($P_\mathrm{c}$) of tropical cyclones (TCs) is an important measure of their intensity. Modern $P_\mathrm{c}$ estimates come from direct and indirect measurements with varying levels of accuracy. Direct measurements include land observations, ship logs, and aircraft reconnaissance, and are often regarded as the ``ground truth'' for studies of TC intensity. However, after routine aircraft reconnaissance in the western North Pacific Ocean was discontinued in 1987, the coverage of direct observations became largely insufficient across global basins. Numerous indirect estimates of $P_\mathrm{c}$ have been developed based on remote sensing observations, which provide broader spatial and temporal coverage. However, substantial uncertainties remain in these estimates.

Modern operational satellite intensity estimates still largely rely on the Dvorak technique, which infers intensity from organized cloud-top patterns in geostationary infrared imagery \citep{dvorak1984tropical, velden2006dvorak}. Other widely used methods include the Advanced Dvorak Technique (ADT) \citep{olander2007advanced} and the Satellite Consensus (SATCON) \citep{velden2020consensus}. Among these remote sensing methods, a physically motivated class of algorithms estimates intensity from satellite-measured TC warm cores, which are especially relevant for $P_\mathrm{c}$ estimation. These methods use passive microwave sounders, such as the Advanced Microwave Sounding Unit (AMSU) and the Advanced Technology Microwave Sounder (ATMS), and derive $P_\mathrm{c}$ from observed brightness temperature ($BT$) anomalies or physically retrieved warm cores \citep{kidder2000satellite, brueske2003satellite, herndon2004upgrades, demuth2006improvement, zhang2019tropical}.

The absolute error of $P_\mathrm{c}$ grows with intensity for most satellite-based methods, making accurate intensity estimation increasingly difficult as storms intensify. \citet{torn2012uncertainty} show that the Dvorak technique's $P_\mathrm{c}$ root-mean-square error (RMSE) grows from \qty{6}{\hecto\pascal} for Tropical Storms to \qty{14}{\hecto\pascal} for Category 4/5 TCs. This trend is particularly significant for methods based on instruments with limited resolution, such as the AMSU \citep{kidder2000satellite} and the Soil Moisture Active Passive (SMAP) radiometer \citep{meissner2017capability}. According to \citet{demuth2006improvement}, the $P_\mathrm{c}$ RMSE of the early AMSU algorithm grows from \qty{6.1}{\hecto\pascal} for Tropical Depressions to \qty{36.3}{\hecto\pascal} for Category 5 TCs. The root cause is that passive microwave sounders cannot sufficiently resolve the inner core of intense tropical cyclones. As tropical cyclones intensify, their inner core scale, characterized by the radius of maximum winds (RMW), generally contracts \citep{Kimball04RMW}. The instantaneous fields of view (IFOVs) of AMSU and ATMS are \qty{48}{\km} and \qty{32}{\km} at nadir for the primary sounding bands, larger than the median RMW of \qty{27.8}{\km} for Category 5 hurricanes \citep{Kimball04RMW}. Therefore, the warm-core signal can be spatially smoothed and contaminated by eyewall convection.

Infrared (IR) imagers and sounders provide an alternative pathway to resolve the warm cores of TCs with mature eyes at higher resolution. The Atmospheric Infrared Sounder (AIRS) has finer resolution (IFOV = \qty{13.5}{\km}) than microwave sounders and has been used in climatological studies of TC warm cores \citep{Wang19AIRS}. However, current IR sounders still fail to resolve TCs with small or partially obscured eyes, as high-level clouds heavily absorb IR emission in the \qtyrange{10}{15}{\um} range \citep{warren2008optical}.

The Moderate Resolution Imaging Spectroradiometer (MODIS) offers unique, long-overlooked capabilities that could revolutionize $P_\mathrm{c}$ estimation for intense TCs. MODIS has six $\mathrm{CO_2}$ absorption bands with \qty{1}{\km} resolution at nadir \citep{esaias1998overview}, sampling air temperatures at various levels of TC eyes. Aboard Aqua and Terra, the MODIS mission spans a 25-year period and observes most tropical regions four times a day, providing an extensive dataset of TC observations. To demonstrate the value of MODIS's high spatial resolution, we develop a $P_\mathrm{c}$ estimation algorithm based on MODIS Thermal Infrared Sounding (MOTIS) and show that it outperforms all existing satellite-based methods for intense (Category 3+) TCs with clear eyes observed. Furthermore, the precision of the estimated TC central pressure ($P_\mathrm{MOTIS}$) does not degrade even for the most intense TCs ($P_\mathrm{c} < 920~\mathrm{hPa}$). We also develop a comprehensive $P_\mathrm{MOTIS}$ dataset for all MODIS passes of mature TC eyes since 2002.

We organize the paper as follows. Section~\ref{sec:data} describes the data used in our framework. Section~\ref{sec:methods} explains the physical basis of the algorithm and the regression parameters. Section~\ref{sec:results} presents our algorithm with its validation performance and demonstrates MOTIS applications for TC cases and reanalysis. Section~\ref{sec:dis} discusses potential improvements and future instruments for Thermal Infrared Sounding (TIS).

\section{Data}\label{sec:data}

To our knowledge, MODIS-based warm-core retrievals have been limited to individual case studies \citep{Durden10Monica} using Level-2 atmospheric profile products \citep{borbas2016modis}, which are downsampled to 5\,km resolution and are not designed for the environment inside TC eyes. Therefore, we estimate $P_\mathrm{c}$ directly using the $BT$ anomaly retrieved from MODIS L1B data, paired with fifth-generation ECMWF reanalysis (ERA5) data for background temperatures.

\subsection{Satellite data}
This study uses L1B calibrated radiances from MODIS Collection 6.1 (C6.1) aboard the Terra and Aqua satellites. The operational products began on 25 February 2000 for Terra and 25 June 2002 for Aqua. For Terra, Bands~33--36 exhibit known mirror side differences before 18 September 2002 because of several configuration changes \citep{chang2024modis_c7_memo}; therefore, we exclude affected Terra observations before this date. We use all remaining observations through 31 December 2025.

MODIS has 36 spectral bands ranging from \qtyrange{0.4}{14.4}{\um}, covering visible (VIS), near-infrared (NIR), shortwave infrared (SWIR), and thermal infrared (TIR) regions. Among the TIR bands, there are six $\mathrm{CO_2}$ absorption bands (Bands~24--25 and Bands~33--36) sensitive to tropospheric temperature. After excluding Band~25 because of its excessive sensitivity to low-level clouds and sunlight and including Bands~27, 31, and 32, a total of eight bands are adopted in this study, with their roles summarized in appendix~B. Figure \ref{figure:wf} shows the weighting functions of the adopted $\mathrm{CO_2}$ absorption bands, representative of the tropospheric levels to which they are sensitive.

\begin{figure}[htbp]
    \centering
	\noindent\includegraphics[width=\columnwidth]{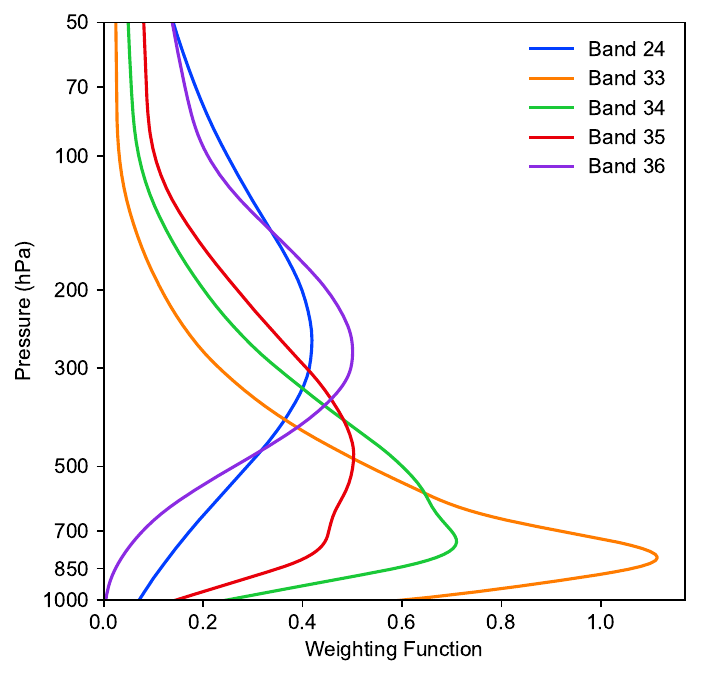}
	\caption{Weighting function of the five Aqua $\mathrm{CO_2}$ absorption bands used in this study, simulated by RTTOV under standard atmosphere and $0\degree$ zenith angle. $\mathrm{CO_2}$ absorption generally strengthens from Band~33 to Band~36, shifting the peak of the weighting function upward.}\label{figure:wf}
\end{figure}

Radiative transfer simulations of satellite observations were performed using RTTOV (Radiative Transfer for TOVS) v13 \citep{saunders2018update}. We use RTTOV to simulate the weighting functions of spectral bands, calculate environmental $BT$, and perform zenith angle adjustment.

\subsection{Reanalysis data}
This study uses the ECMWF fifth-generation reanalysis ERA5 \citep{hersbach2020era5}, obtained from the Copernicus Climate Data Store, as the TC environmental background. Each MODIS pass is matched with the closest hourly ERA5 fields on a $0.25^\circ\times0.25^\circ$ grid. The air temperature on pressure levels \citep{cds_era5_pl} and single-level fields of mean sea-level pressure \citep{cds_era5_sl} are used.

\subsection{TC pressure data}

\begin{figure}[htbp]
	\noindent\includegraphics[width=\columnwidth]{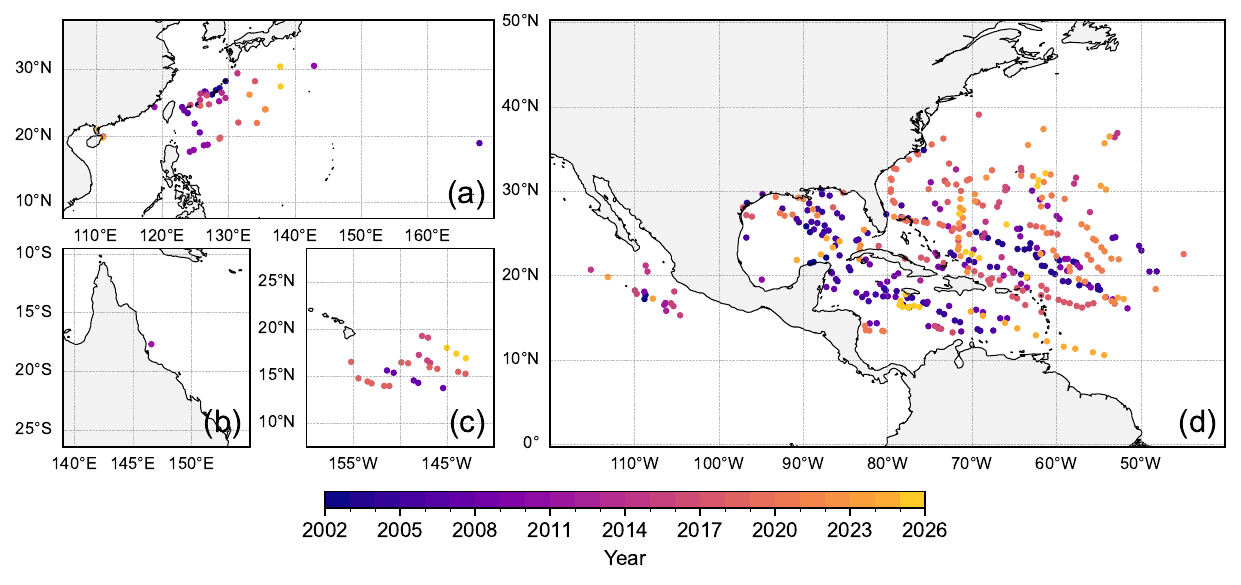}
	\caption{Locations and years of the TC pressure samples. Panels correspond to the (a) western North Pacific, (b) southwest Pacific, (c) central North Pacific, and (d) Atlantic and eastern North Pacific basins.}
    \label{Obs map}
\end{figure}

\begin{table}[htbp]
    \centering
    \footnotesize
    \caption{Number of MODIS passes in each ocean basin.}
    \begin{tabular}{m{2.5cm}<{\centering}m{0.6cm}<{\centering}m{1.4cm}<{\centering}m{1.7cm}<{\centering}}
        \toprule[1pt]
        Basin & Code & Passes with observation & Passes with no observation \\
        \midrule
        Atlantic & AL & 381 & 236 \\
        Eastern North Pacific & EP & 17 & 373 \\
        Central North Pacific & CP & 25 & 89 \\
        Western North Pacific & WP & 43 & 1354 \\
        North Indian & IO & 0 & 62 \\
        Southwest Indian & SI & 0 & 436 \\
        Australian & AU & 1 & 139 \\
        Southwest Pacific & SP & 0 & 132 \\
        Total ($N=3288$) & - & 467 & 2821 \\
        \bottomrule[1pt]
    \end{tabular}
    \label{tab:basin list}
\end{table}

Table~\ref{tab:basin list} summarizes the number of MODIS passes obtained in each basin. From all MODIS passes ($N=3288$) with distinct TC eyes, we match 467 MODIS passes with TC pressure observations to develop MOTIS, as shown in Figure \ref{Obs map}. Most observations were obtained from aircraft reconnaissance missions. For missions in AL ($N=381$), EP ($N=17$), and CP ($N=25$), data were obtained from the Hurricane Research Division (HRD) of the Atlantic Oceanographic and Meteorological Laboratory (AOML), while data from WP missions were obtained from both HRD and the Tropical Cyclones--Pacific Asian Research Campaign for the Improvement of Intensity Estimations/Forecasts (T-PARCII) \citep{tsuboki2017tropical}.

To improve coverage in basins with limited aircraft reconnaissance, pressure data from non-aircraft sources are also included. Drifting buoy observations were provided by the National Oceanographic Data Center (NODC). Land observations were gathered from Global Synoptic Surface Observations and the Japan Meteorological Agency (JMA). Wave glider observations for Typhoon Danas (2013) were obtained from \citet{mitarai2016wave}. Additional radar-based estimates in the western North Pacific were extracted from \citet{shimada2016evaluation}, \citet{shimada2018doppler}, and \citet{ryukyus}.

For MODIS passes that were temporally offset from the observations, the pressure at the pass was interpolated from existing observations or extrapolated based on trends shown by the Regional Specialized Meteorological Centre (RSMC) or Tropical Cyclone Warning Centre (TCWC) Best Track data retrieved from \citet{knapp2010ibtracs,gahtan2024ibtracs}, together with qualitative structural changes captured by geostationary satellites. Passes were included if (i) the closest observation was within 3 hours, or (ii) the pass was between a pair of observations no more than 12 hours apart and Best Track TC central pressure ($P_\mathrm{best}$) did not show a reversal in pressure trend.

Best Track data were not used as the observation set because of regional differences in Best Track practices: aircraft reconnaissance data are directly assimilated in the Best Track of RSMC Miami (National Hurricane Center), but most other RSMC/TCWCs primarily rely on satellite intensity analyses and wind-pressure relationships, with surface observations serving mainly as supplementary constraints \citep{velden2012first}.

\section{Methodology}\label{sec:methods}

\subsection{Physical basis of MOTIS}\label{ssec:basis}

Aircraft observations first showed that intense TCs can contain pronounced warm-core temperature anomalies in the eye \citep{jordan1957estimating}, and later microwave-sounder studies demonstrated that satellite-observed warm-core anomalies are statistically related to $P_\mathrm{c}$ \citep{kidder2000satellite}. This relationship can be understood from hydrostatic balance.

The hypsometric equation \citep{wallace2006atmospheric}, based on the assumption of vertical hydrostatic equilibrium, links sea-level pressure in the TC eye to the mean virtual temperature ($\bar{T}_v$) of the overlying air column. For an idealized air column of height $h$ in an isothermal environment with temperature $T_0$, a small uniform warming $\Delta T_v \ll T_0$ gives the following pressure deficit:
\begin{equation}
\begin{split}
    \Delta P &= P_{\mathrm{env}} - P_\mathrm{c} \\
    &= P_{\mathrm{env}} \left(1-e^{-h\Delta T_v\frac{g}{RT_0(T_0+\Delta T_v)}}\right) \\
    &\approx P_{\mathrm{env}} (\Delta T_v h) \frac{g}{R {T_0}^2}.
\end{split}
\end{equation}
Thus, for a nearly fixed environmental pressure ($P_{\mathrm{env}}$), the TC pressure deficit ($\Delta P$) is proportional, to first order, to the product of the average temperature anomaly and the total height of the perturbed air column.

\subsection{Temperature anomaly}

Because of the limited number of channels and possible cloud absorption, retrieval of the actual temperature anomaly ($\Delta T$) involves significant uncertainties. Instead, we predict $P_\mathrm{c}$ directly from the brightness temperature anomaly ($\Delta BT$), defined as the difference between the highest $BT$ in the eye and the environmental $BT$ ($BT_\mathrm{env}$), where the eye region is found using the approach of \citet{tsukada2023strong} with minor changes. Since the absorption strength of $\mathrm{CO_2}$ increases with temperature, $\Delta BT$ is lower than $\Delta T$ under clear-sky conditions, roughly satisfying $\Delta BT\sim0.5\Delta T$ according to RTTOV simulations. The low amplitude of $\Delta BT$ also implies that more cautious preprocessing is required, as detailed in appendix~C.

\begin{figure}[htbp]
    \centering
	\noindent\includegraphics[width=\columnwidth]{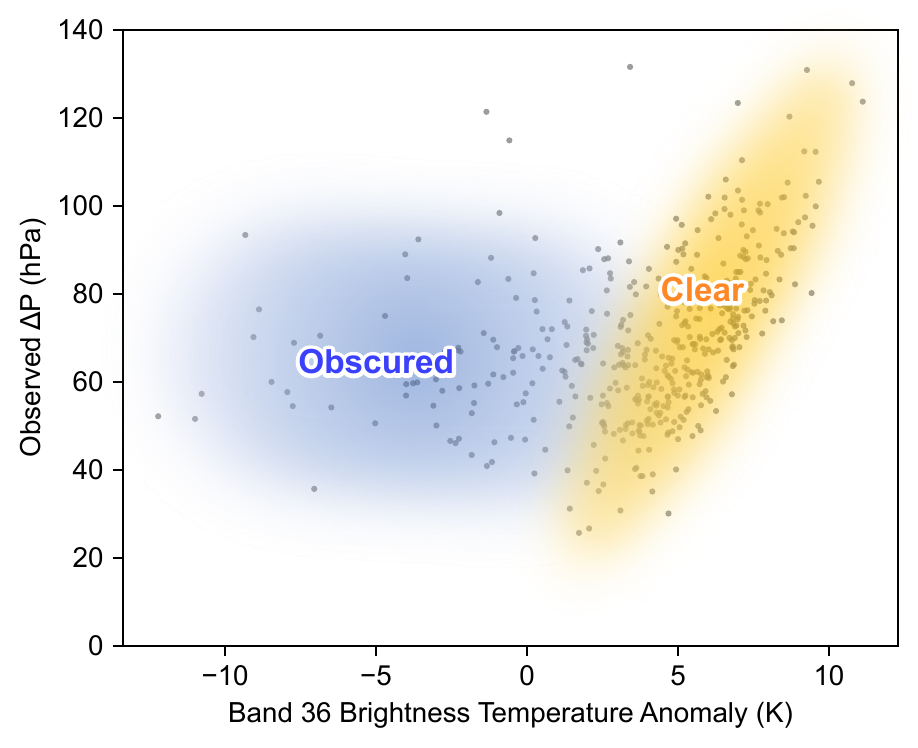}
	\caption{Scatterplot of observed $\Delta P$ against $\Delta T_{\mathrm{B36}}$. The ``clear'' region represents samples with generally clear eyes, where $\Delta P$ is approximately linear in $\Delta T_{\mathrm{B36}}$, with some scatter caused by differences in vertical warm-core structure. The ``obscured'' region has much lower $\Delta T_{\mathrm{B36}}$ than TCs at the same intensity, indicating significant cloud obscuration.}
    \label{Tbd36 vs dp}
\end{figure}

In practice, the correspondence between $\Delta BT$ and the actual $\Delta T$ profile or $\Delta P$ is also influenced by various factors, including the vertical warm-core structure, spatial sampling, and absorption by $\mathrm{H_2O}$ and clouds. Figure \ref{Tbd36 vs dp} demonstrates the effect of absorption: samples in the ``obscured'' region have cloud absorption in the eye, showing lower $\Delta BT$ than TCs with clear eyes. Similar to the approach in the AMSU algorithm \citep{demuth2006improvement}, we incorporate auxiliary parameters detailed in appendix~D to obtain a more precise estimate of the overall temperature perturbation and the TC central pressure.

\begin{figure}[htbp]
    \centering
	\noindent\includegraphics[width=\columnwidth]{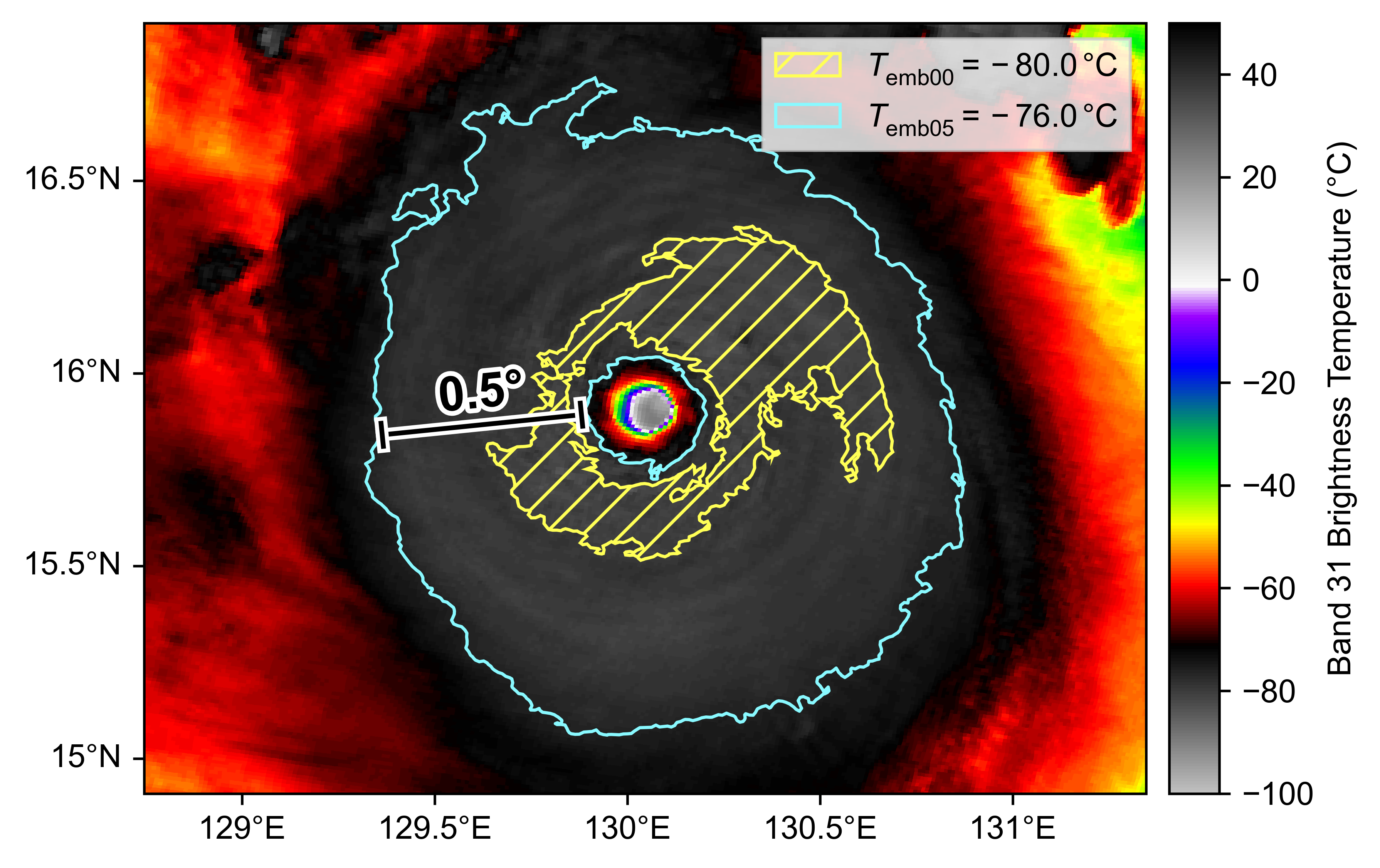}
	\caption{Aqua Band~31 imagery of Typhoon Goni (2020) at 16:55 UTC on Oct 30. The contours and hatching show the coldest connecting regions with minimum widths of 0.0 and 0.5 degrees latitude, corresponding to $T_\mathrm{emb00}$ and $T_\mathrm{emb05}$.}\label{figure:cdotemp}
\end{figure}

\subsection{Height of warm core}\label{sssec:height}

MODIS cannot directly determine the vertical range of the TC warm core, as the weighting functions of all $\mathrm{CO_2}$ absorption bands peak well below the tropical tropopause ($\sim$\qty{100}{\hecto\pascal}). To address this limitation, we use the Band~31 $BT$ of the TC's central dense overcast (CDO) as an approximation of the warm-core top height, under the assumption that the height of persistent eyewall convection is representative of the height of the warm core.

To estimate the CDO temperature, we first interpolated $BT$ data onto polar coordinates centered on the eye. We then measured the coldest connecting ring meeting width requirements ranging from 0 to 1 degree of latitude, in 0.1 degree intervals. These embedding temperatures are denoted as $T_\mathrm{emb00}$, $T_\mathrm{emb01}$, ..., and $T_\mathrm{emb10}$. Figure~\ref{figure:cdotemp} illustrates this procedure for Typhoon Goni (2020).

After experimenting with different width requirements, we found a weighted average of $T_\mathrm{emb00}$ and $T_\mathrm{emb05}$ to be most representative of the observed TC $\Delta P$ and adopted it in the regression as $T_\mathrm{CDO}$. This choice of CDO ring widths also coincidentally matches the IR eye pattern in the Dvorak technique \citep{dvorak1984tropical, velden2006dvorak}. $T_\mathrm{CDO}$ is calculated as below, followed by a zenith angle adjustment (see appendix~C, section~\ref{ssec:zenith}):
\begin{equation}
    T_\mathrm{CDO}=\frac{T_\mathrm{emb00}+0.7\times T_\mathrm{emb05}}{1.7} + \Delta T_\mathrm{CDO, adj.}(\theta)
\end{equation}

$T_\mathrm{CDO}$ was then capped to be no warmer than \qty{-50}{\degreeCelsius}, because TCs with warm CDO still have a significant Band~36 warm anomaly, suggesting that the height of the warm-core top does not substantially decrease below a threshold.

Since $T_\mathrm{CDO}$ is a proxy for the total height of the TC warm core, the difference between $T_\mathrm{CDO}$ and a prespecified $T_\mathrm{base}=35\degree \mathrm{C}$ was calculated. It was then normalized such that the mean was around 1, using an estimated average $T_\mathrm{CDO}$ of \qty{-65}{\degreeCelsius}:

\begin{equation}
    T_\mathrm{CDO,norm.}=\frac{T_\mathrm{base}-T_\mathrm{CDO}}{T_\mathrm{base}-(-65\degree \mathrm{C})}
\end{equation}

\subsection{Environmental temperature and pressure}

Since it is difficult to accurately retrieve $BT_\mathrm{env}$ directly from satellite observations because of cloud absorption, we estimate the $\Delta BT$ of Band~24 and Bands~33--36 by applying RTTOV simulations to the ERA5 background temperature profile, defined as the median profile within a TC-centered annulus spanning $r=12^\circ$--$13^\circ$ latitude, similar to the annulus size choice in \citet{frank1977structure}. The median is preferred over the mean because the latter is more easily biased for high-latitude TCs, where the annulus partly samples temperate regions. In RTTOV simulations, a linear formula was used to account for the rising $\mathrm{CO_2}$ concentration with time, which increases the atmospheric opacity and leads to a gradual decrease of $BT$.

We use the pressure of the outermost closed isobar (POCI) of TCs as $P_{\mathrm{env}}$, obtained by calculating isobars on the ERA5 mean sea-level pressure (MSLP) field at \qty{0.1}{\hecto\pascal} intervals. Among the $P_{\mathrm{env}}$ definitions tested, POCI yielded more accurate $P_\mathrm{c}$ estimates than the mean/median MSLP computed at any fixed radius from the TC center, as the latter may be affected by MSLP inaccuracies over nearby landmasses.

\section{Results}\label{sec:results}

\subsection{Model performance}\label{ssec:model}

\begin{table*}[t]
\centering
\footnotesize
\renewcommand{\arraystretch}{1.5}
\begin{tabular}{m{2.8cm}  m{2.8cm}  m{2.6cm}  m{1.5cm}  m{1.5cm}}
\topline
Parameter $X_i$ & Definition & Role & Importance & Coefficient $\beta_i$ ($N_x=10$)\\
\midline
Band~24 T & $\Delta T_\mathrm{B24} $ & Whole layer warm core & High & 3.4313\\
Band~36 T & $\Delta T_\mathrm{B36} $ & Upper warm core & Very High & 5.6742\\
Upper $BT$ difference & $\frac{\Delta T_\mathrm{B36} - \Delta T_\mathrm{B35}}{\mathrm{max}(\Delta T_\mathrm{B36} + \Delta T_\mathrm{B35},9)} $ & Distribution of upper warm core & Medium & 42.7566\\
Mid-level $BT$ difference & $\frac{\Delta T_\mathrm{B35} - \Delta T_\mathrm{B34}}{\mathrm{max}(\Delta T_\mathrm{B35} + \Delta T_\mathrm{B34},9)} $ & Distribution of mid-level warm core & Low & -28.7243\\
Band~27 T & $\Delta T_\mathrm{B27} $ & Excess $\mathrm{H_2O}$ absorption & Low & -0.0785\\
Eye size & $\frac{1}{\sqrt{\mathrm{max}(8, n_\mathrm{B35\ 1K\ pix.})}} $ & Undersampling  & Very High  & 61.6813\\
Bands~31--32 difference & $T_\mathrm{B31}-T_\mathrm{B32} $ & Cirrus absorption & Medium & -1.3833\\
Truncated Band~33 T & ${\Delta T}_\mathrm{B33, trunc.} $ & Obscured eye $BT$ bias & Very High & 2.5781\\
Instrument bias (Aqua) & $ \begin{cases} T_{\mathrm{rfl.}} & \mbox{if Aqua daytime} \\ 0 & \mbox{otherwise} \end{cases}  $ & Aqua Band~24 daytime reflection & Medium & -1.0099\\
Instrument bias (Terra) & $ \begin{cases} {\Delta T}_\mathrm{B34} & \mbox{if Terra} \\ 0 & \mbox{if Aqua} \end{cases}\rule[-3.5ex]{0pt}{8ex}  $ & Terra overall bias & Medium & -0.6308\\
\midline
CDO Temperature & $T_{\mathrm{CDO,norm}}$ & Warm core top height & Very High & - \\
Regression Constant & $\beta_0$ & - & - & 12.8 \\
Pressure Deficit &
\multicolumn{4}{c}{\rule[-3.5ex]{0pt}{8ex}$\displaystyle
\Delta P=\beta_0+T_{\mathrm{CDO,norm}}\sum_{i=1}^{N_x}\beta_i X_i
$} \\
\botline
\end{tabular}
\caption{Parameters and components of the main MOTIS algorithm; also see section~\protect\ref{sec:methods} and appendix~D.}
\label{tab:parameters}
\end{table*}

\begin{table}[t]
\begin{center}
\footnotesize
\caption{Performance of the main and simplified algorithms.}
\label{tab:performance}
\renewcommand{\arraystretch}{1.3}
\begin{tabular}{m{3cm}<{\centering}m{1cm}<{\centering}m{0.9cm}<{\centering}m{0.9cm}<{\centering}}
\toprule[1pt]
Algorithm & $r^2$ & MAE (\unit{\hecto\pascal}) & RMSE (\unit{\hecto\pascal}) \\
\midrule
\textbf{Main} ($N_x=10$) & \textbf{0.945}$^{\rm a}$ (0.847)  & \textbf{3.4} (5.0) & \textbf{4.3} (6.9) \\
Exclude low importance ($N_x=8$) & 0.944 (0.841)  & 3.5 (5.1) & 4.3 (7.0) \\
Exclude low and medium ($N_x=4$) & 0.913 (0.789)  & 4.3 (5.8) & 5.5 (8.1) \\
\bottomrule[1pt]
\end{tabular}
\end{center}

\footnotesize{$^{\rm a}$ Values outside parentheses are for good-eye samples; values in parentheses are for all samples. Good-eye criteria are given in Table~\ref{tab:eye_type}.}


\end{table}

Using the physical predictors described in section~\ref{sec:methods} and the auxiliary correction terms detailed in appendix~D, we constructed our main 10-parameter algorithm using multiple linear regression. Descriptions of the parameters and their qualitative importance in the model, determined \textit{a posteriori}, are shown in Table \ref{tab:parameters}. We measured model performance with $r^2$, mean absolute error (MAE), and RMSE. The regression coefficients and performance of our main algorithm (number of parameters $N_x=10$) are shown in Table \ref{tab:performance}. Two selected reduced-parameter algorithms ($N_x=8,4$) are also included for comparison, illustrating the measurable improvement provided by the medium-importance parameters and the marginal improvement provided by the low-importance parameters.

\begin{figure}[htbp]
    \centering
	\noindent\includegraphics[width=\columnwidth]{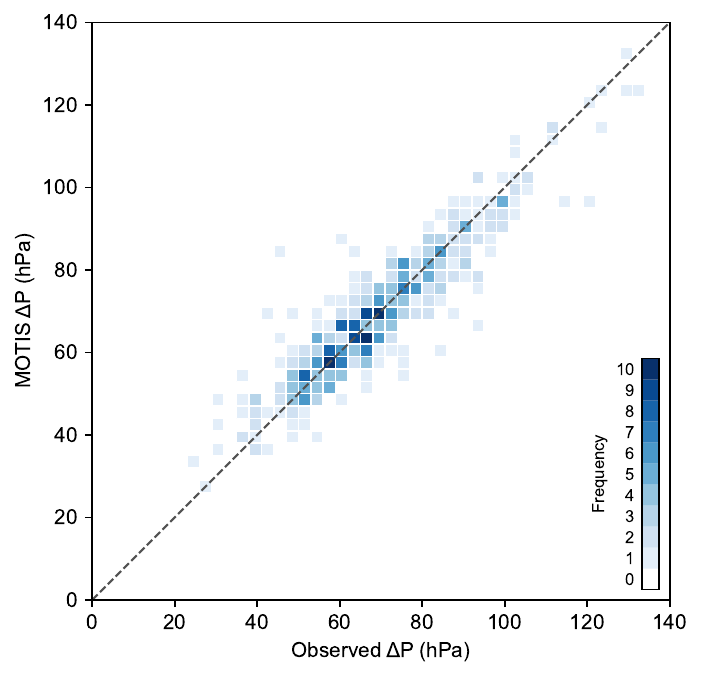}
	\caption{2D histogram of modeled versus observed TC pressure deficit for the full observation sample. Bins are \qty{3}{\hecto\pascal} wide.}
    \label{Model vs Obs}
\end{figure}

\begin{figure*}[htbp]
    \centering
	\noindent\includegraphics[width=1.9\columnwidth]{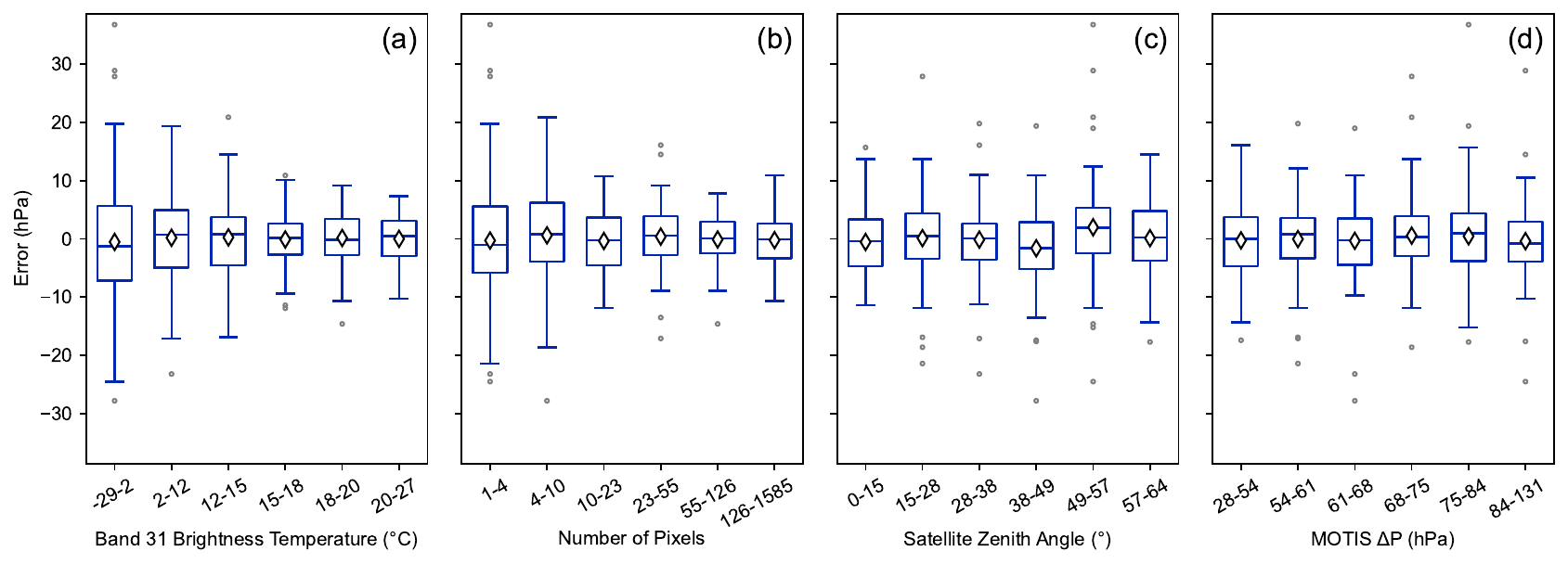}
	\caption{Error distribution of MOTIS estimates shown in box plots with equal-frequency bins of (a) Band~31 $BT$, (b) $n_{\mathrm{pix}}$, (c) satellite zenith angle, and (d) Model $\Delta P$. MOTIS's precision increases with increasing $BT$ and $n_{\mathrm{pix}}$, while showing no significant trend with zenith angle or TC intensity.}
    \label{fig:error distribution}
\end{figure*}

Figure \ref{Model vs Obs} compares the MOTIS pressure deficit with observations. The mean of the observed pressure deficit ($\Delta P_\mathrm{obs}$) is \qty{69.3}{\hecto\pascal}, which approximately corresponds to a Category 4 TC with $P_\mathrm{c}$ $\approx$ \qty{942}{\hecto\pascal} (mean POCI = \qty{1011}{\hecto\pascal}).

A closer examination shows that MOTIS's precision depends strongly on whether the eye is clear or obscured. As shown in Figure \ref{fig:error distribution} (a), the uncertainty decreases significantly with increasing eye temperature $T_\mathrm{B31}$. The uncertainty also decreases with the number of eye pixels ($n_{\mathrm{pix}}$, see appendix~D, section~\ref{ssec:pixel}), as shown in Figure \ref{fig:error distribution} (b).

In contrast, Figure \ref{fig:error distribution} (c) and (d) show that the error has no clear trend with satellite zenith angle or TC intensity. Although many existing methods have larger errors for high-intensity TCs, MOTIS demonstrates stable performance even for the highest intensities ($\Delta P_\mathrm{obs}$ = 84--131\,\unit{\hecto\pascal}, $P_\mathrm{obs}$ $\approx$ 927--880\,\unit{\hecto\pascal}).

\begin{table}
    \centering
    \caption{Classification of eye clarity.}
    \label{tab:eye_type}
    \renewcommand{\arraystretch}{1.3}
    \begin{tabular}{c c}
    \toprule[1pt]
    Type & Criteria \\
    \hline
    Good & $T_{\mathrm{B31}} \ge 16^\circ\mathrm{C} \text{ and } \Delta T_{\mathrm{B33}} \ge -1\,\mathrm{K} \text{ and } n_{\mathrm{pix}} \ge 4$ \\
    Fair & $T_{\mathrm{B31}} \ge 2^\circ\mathrm{C} \text{ and } \Delta T_{\mathrm{B33}} \ge -13\,\mathrm{K}$ \\
    Poor & Others \\
    \bottomrule[1pt]
    \end{tabular}
\end{table}

\begin{figure*}[htbp]
    \centering
	\noindent\includegraphics[width=1.6\columnwidth]{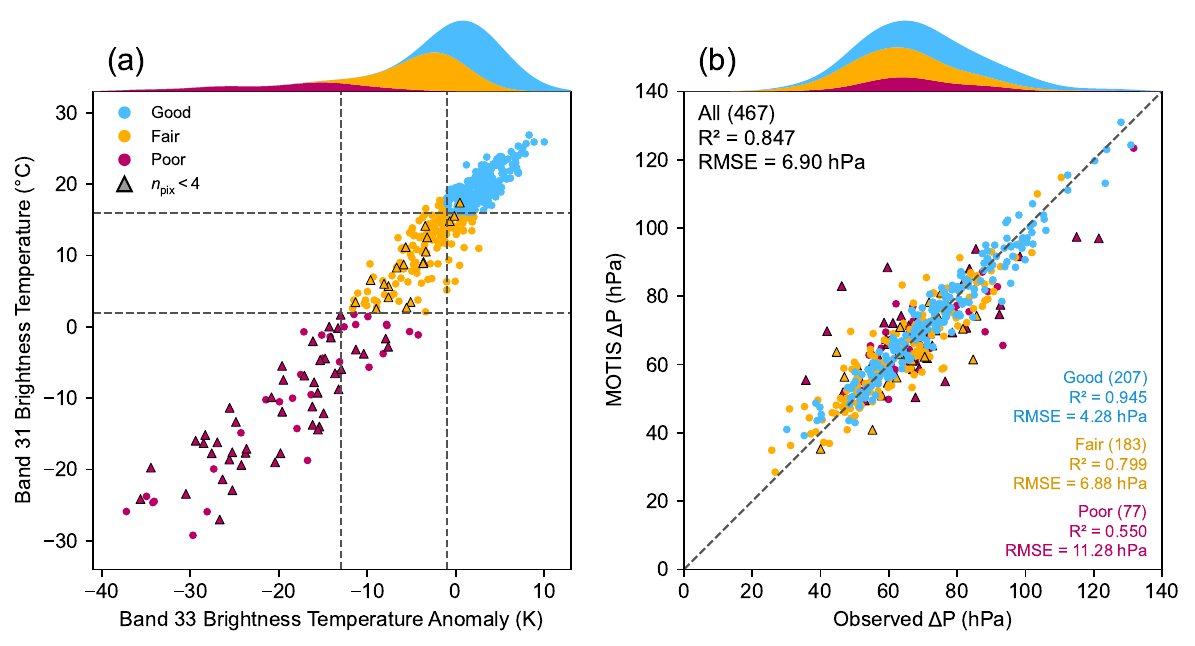}
	\caption{Classification of eye clarity and model performance within each group. Panel (a) shows the groups' distribution with $T_\mathrm{B31}$ and $\Delta T_\mathrm{B33}$. Samples with $n_{\mathrm{pix}}<4$ are marked with triangles instead of circles, and the threshold temperatures are labeled with dotted lines. Panel (b) shows the scatter of $\Delta P_\mathrm{MOTIS}$ for each TC group against $\Delta P_\mathrm{obs}$. Consistent with Figure \protect\ref{Model vs Obs} and Figure \protect\ref{fig:error distribution}, the most extreme outliers disproportionately come from the poor-eye group. The frequency distributions of $\Delta T_\mathrm{B33}$ and $\Delta P_\mathrm{obs}$ are shown above panels (a) and (b), respectively.}
    \label{fig:quality_group}
\end{figure*}

To further examine the effect of eye clarity on MOTIS's precision, we flagged the MODIS passes based on the criteria in Table \ref{tab:eye_type}. The good-eye group ($N=207$) exhibits significantly improved precision, with $r^2=0.945$ and $\mathrm{RMSE}=$ \qty{4.3}{\hecto\pascal}. Figure \ref{fig:quality_group} presents the eye clarity classification and the error distribution of each group. Good-eye samples become more common at higher intensities, with $\overline{\Delta P_\mathrm{c}}=$ \qty{74.4}{\hecto\pascal} and $\overline{P_\mathrm{c}}=$ \qty{937}{\hecto\pascal}. This may partly explain why MOTIS errors remain small for the most intense TCs (Figure~\ref{fig:error distribution}).

The advantage of MOTIS is most pronounced for high-intensity TCs. For comparison, the Dvorak technique has a reported $P_\mathrm{c}$ RMSE of \qty{14}{\hecto\pascal} for Category~4/5 TCs \citep{torn2012uncertainty}, and the CIRA AMSU algorithm has an RMSE of \qty{36.3}{\hecto\pascal} for Category~5 TCs \citep{demuth2006improvement}. An updated internal validation of the CIMSS AMSU algorithm \citep{herndon2004upgrades} gives a Category~5 RMSE of \qty{16}{\hecto\pascal}. For the good-eye MOTIS estimates, the RMSE remains \qty{4.4}{\hecto\pascal} for $P_\mathrm{obs}<\qty{945}{\hecto\pascal}$, \qty{5.1}{\hecto\pascal} for $P_\mathrm{obs}<\qty{925}{\hecto\pascal}$, and \qty{4.9}{\hecto\pascal} for $P_\mathrm{obs}<\qty{900}{\hecto\pascal}$, roughly corresponding to Category~4/5 TCs, Category~5 TCs, and the top seven lowest pressures in Atlantic on record.

The RMSE and bias have no significant long-term or seasonal trend. When the dataset is split into the training dataset (2003--2018) and verification dataset (2019--2025), the RMSE of the verification dataset remains \qty{6.9}{\hecto\pascal} for the full sample and \qty{4.3}{\hecto\pascal} for the good-eye group, indicating minimal overfitting and robust generalization over the multi-decadal record. For diurnal variations, we note that MOTIS's precision is worse for daytime poor-eye samples, presumably because Band~24's sunlight contribution (see appendix~D, section~\ref{ssec:bias}) is not fully corrected.

MOTIS's performance is generally consistent across basins. For the good-eye group, the RMSE is slightly larger in the western North Pacific (\qty{4.7}{\hecto\pascal}) than the average (\qty{4.2}{\hecto\pascal}), which we attribute to larger uncertainties in non-aircraft observations rather than an intrinsic trait of MOTIS. There is limited evidence of basin-specific bias, with estimates \qty{1.3}{\hecto\pascal} too weak in the western North Pacific and \qty{1.5}{\hecto\pascal} too strong in the eastern and central North Pacific. This bias is further discussed in section~\ref{sec:dis}\ref{ssec:systematic}.

\subsection{Case applications}

\begin{figure*}[htbp]
    \centering
	\noindent\includegraphics[width=1.9\columnwidth]{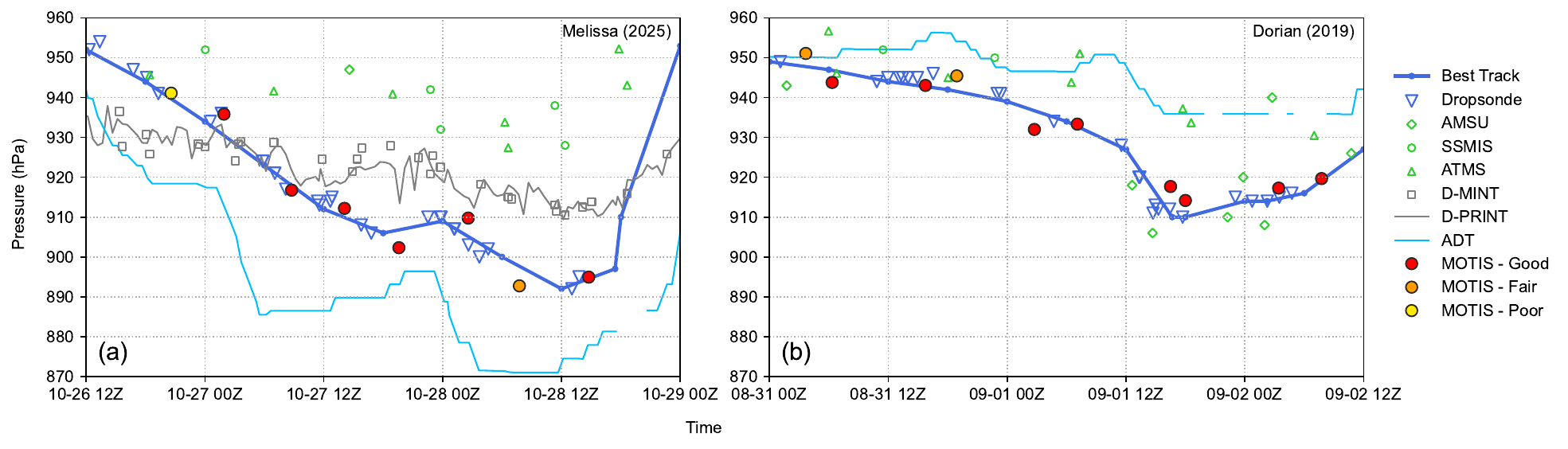}
\caption{Central-pressure estimates for (a) Hurricane Melissa (2025) and (b) Hurricane Dorian (2019). MOTIS estimates are grouped by eye-clarity class and compared with Best Track, dropsonde observations, CIMSS microwave sounder estimates (AMSU, SSMIS, and ATMS), CNN-based satellite estimates (D-MINT and D-PRINT; not available for Dorian), and ADT. Times are in UTC.}
    \label{case}
\end{figure*}

To illustrate the application of MOTIS in individual TC cases, we compare MOTIS estimates with Best Track, dropsonde observations, and numerous SATCON-member satellite estimates. Figure \ref{case} shows Hurricane Melissa (2025) and Hurricane Dorian (2019), two extremely intense landfalling TCs with dense reconnaissance observations. All of the MOTIS estimates are within \qty{5}{\hecto\pascal} from the reconnaissance-supported Best Track, with the exception of 01:55 UTC on Oct 27 (\qty{5.3}{\hecto\pascal}) for Melissa. Existing satellite-based estimates are far less accurate, especially near peak intensity. At Melissa's peak intensity (12:00 UTC on Oct 28), ADT is \qty{19}{\hecto\pascal} too strong, D-MINT and D-PRINT are $\sim$\qty{19}{\hecto\pascal} too weak, and SSMIS is $\sim$\qty{40}{\hecto\pascal} too weak. By contrast, the nearest MOTIS estimate \qty{3}{\hour} later is only \qty{0.5}{\hecto\pascal} weaker than the Best Track value. This suggests that MOTIS could be incorporated into SATCON as an important future member, potentially improving the accuracy of SATCON's multi-source consensus for high-intensity TCs.

\subsection{Comparison with Best Track} \label{ssec:reanalysis}

The 25-year MODIS mission provides an extensive dataset of warm core observations and $P_\mathrm{c}$ estimates for high-intensity TCs. We obtained 3288 MODIS passes of mature TCs with distinct eye features, among which 1082/1408/798 are classified as good/fair/poor eyes. In particular, 70\% of all Atlantic major hurricanes during the period have had at least one MODIS pass with a ``good'' eye, despite MODIS's restricted temporal coverage (maximum of four passes per day). This fraction increases to 83\% and 89\% after including fair- and poor-eye passes. The prevalence of good-eye passes increases to 77\% when only considering Category 4\&5 hurricanes, and to 90\% for Category 5 hurricanes alone.

Here, we demonstrate the potential of MOTIS estimates to improve Best Track quality and to identify basin-specific bias in Best Track TC central pressure ($P_\mathrm{best}$) among RSMC/TCWCs.

\begin{figure}[htbp]
    \centering
	\noindent\includegraphics[width=\columnwidth]{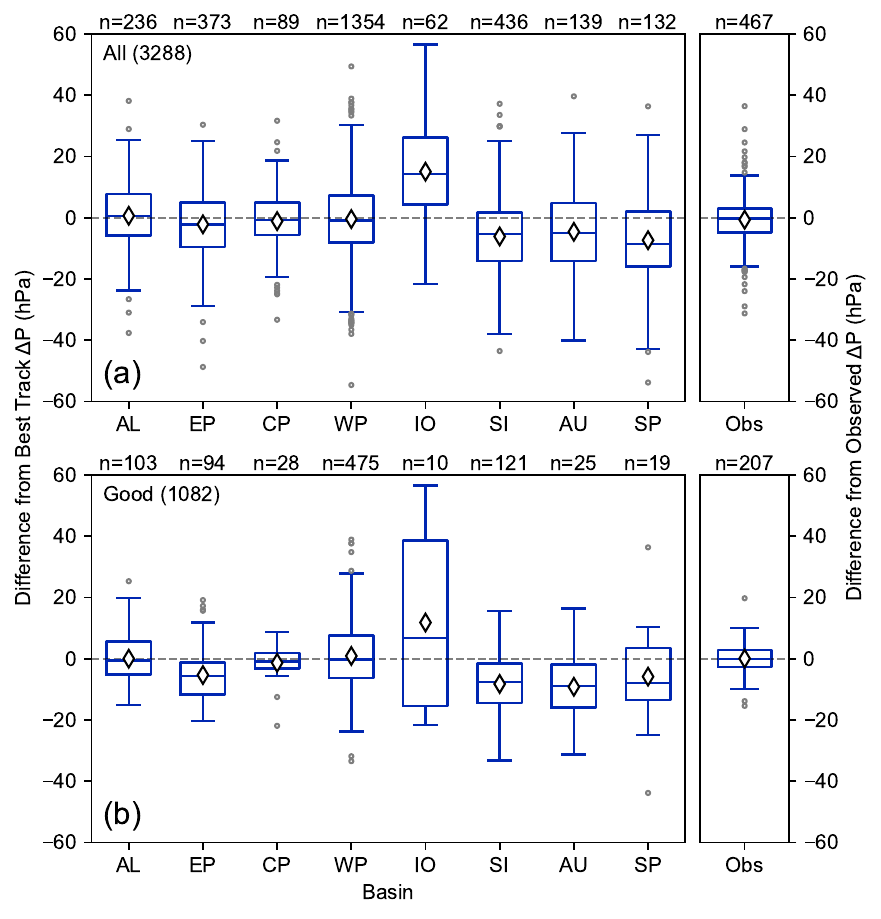}
	\caption{Box plots comparing $P_\mathrm{MOTIS}$ and $P_\mathrm{best}$ or pressure observations across basins for (a) all samples and (b) good-eye samples. Basin codes are defined in Table \protect\ref{tab:basin list}. The left panels use non-observation Best Track, and the right panels use pressure observations, representing MOTIS's accuracy against ground truth. Larger values indicate stronger MOTIS estimates.}
    \label{bst}
\end{figure}

Figure \ref{bst} shows the difference between $P_\mathrm{best}$ and $P_\mathrm{MOTIS}$ by basin in the absence of pressure observations. The error distribution of $P_\mathrm{MOTIS}$ against $P_\mathrm{obs}$, representing MOTIS's accuracy against ground truth, is shown alongside as a reference.
The RMS difference against non-observation $P_\mathrm{best}$ is much larger (All: \qty{12.9}{\hecto\pascal}; Good: \qty{11.4}{\hecto\pascal}) than the RMSE against pressure observations (All: \qty{6.9}{\hecto\pascal}; Good: \qty{4.3}{\hecto\pascal}) and depends strongly on RSMC/TCWC in each basin.

Evidence suggests that the large RMS difference arises primarily from uncertainties in the Best Track data, rather than from generalization error in the MOTIS framework. First, the $P_\mathrm{MOTIS}$ RMSE shows no visible degradation when split into the training dataset (2003--2018) and verification dataset (2019--2025), and it shows consistent accuracy across the four major basins with observation samples. Second, the RMS difference is numerically consistent with previous studies of Best Track uncertainties. \citet{landsea2013atlantic} suggest that Atlantic (AL) $P_\mathrm{best}$ without aircraft observation has an RMSE of \qty{7.7}{\hecto\pascal} for Category 1--2 TCs and \qty{9.5}{\hecto\pascal} for Category 3+ TCs. This is close to the calculated RMS differences of \qty{7.7}{\hecto\pascal} (Good) and \qty{10.5}{\hecto\pascal} (All) in AL, the latter presumably affected by MOTIS's higher uncertainty for poor-eye samples.
As shown in Figure \ref{bst}, other basins show a comparable or larger RMS difference than AL, with the exception of the central North Pacific (CP), where the smaller sample group consists mostly of annular hurricanes with stable intensity \citep{knaff2003annular}. The North Indian Ocean (IO) shows the largest RMS difference of \qty{30.3}{\hecto\pascal} for the good-eye samples.
For good-eye samples, we conclude that MOTIS can substantially reduce the Atlantic Best Track intensity uncertainty by about 50\% in the absence of aircraft observations, and potentially by even more in other basins.

In addition, systematic biases are observed in the Best Track records of several RSMC/TCWCs. In the full sample, $P_\mathrm{MOTIS}$ is \qty{12}{\hecto\pascal} stronger than $P_\mathrm{best}$ in IO and \qtyrange{6}{8}{\hecto\pascal} weaker in SI/AU/SP. The bias in IO can be explained by the lack of high-quality geostationary observations \citep{knapp2008scientific}, which makes it difficult to fully resolve the eye of high-intensity TCs in the operational Dvorak/ADT workflow (see \citet{velden2006dvorak, olander2019advanced}). Since these tools depend strongly on the resolved eye temperature, insufficient instrument resolution can cause significant weak biases in $P_\mathrm{best}$. On the other hand, RSMC/TCWCs in SI/AU/SP historically adopted the Atkinson--Holliday \citep{atkinson1977tropical} or Crane wind-pressure relationship \citep{courtney2009adapting}, which sometimes yield much lower $P_\mathrm{c}$ than approaches such as the modern KZC relation \citep{knaff2007reexamination}. Hence, $P_\mathrm{best}$ may appear stronger in SI/AU/SP.

\begin{figure}[htbp]
    \centering
	\noindent\includegraphics[width=\columnwidth]{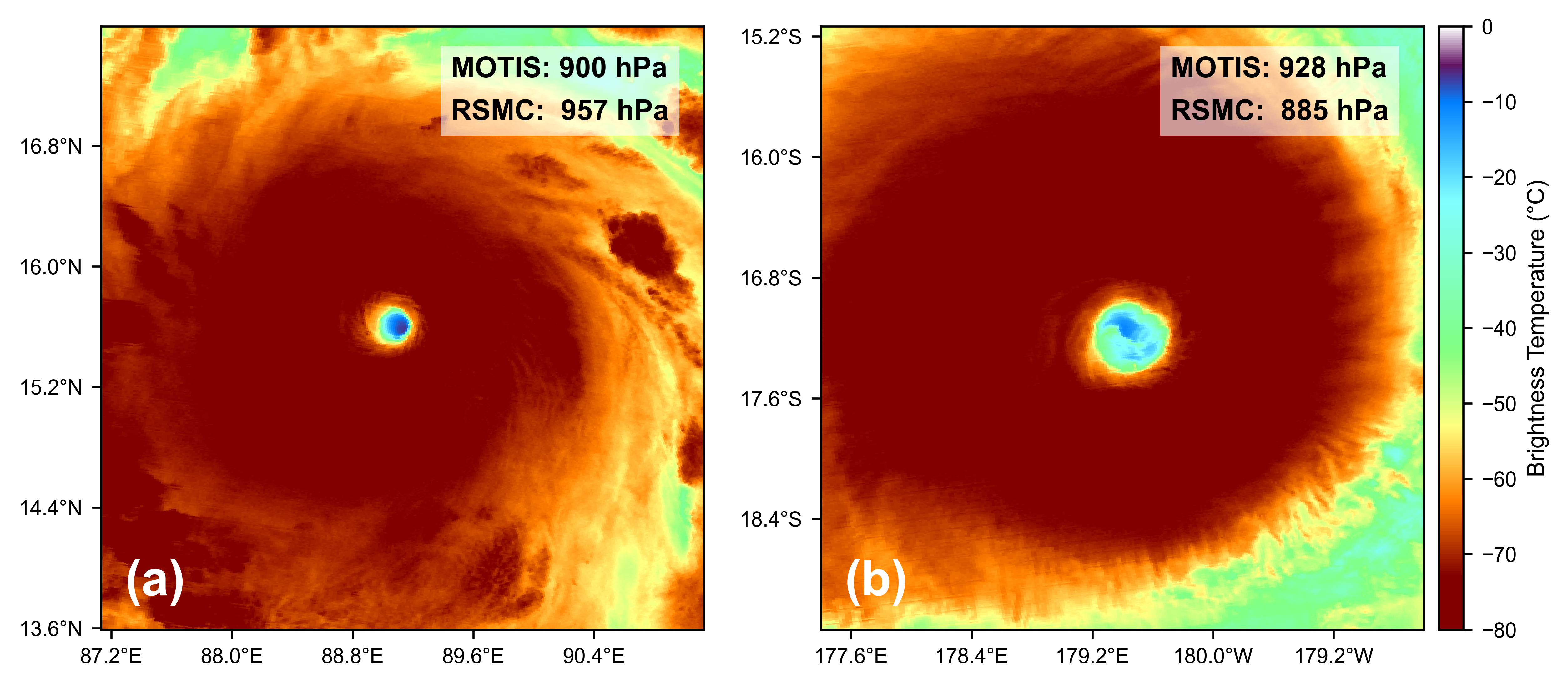}
	\caption{The Band~34 imagery of Cyclone Phailin (2013) at 19:30 UTC on Oct 10 and Cyclone Winston (2016) at 01:30 UTC on Feb 20.}
    \label{phailin}
\end{figure}

Figure \ref{phailin} presents two cases with large discrepancies between $P_\mathrm{best}$ and $P_\mathrm{MOTIS}$. Cyclone Phailin (2013) is \qty{57}{\hecto\pascal} stronger in MOTIS than in the interpolated Best Track from RSMC New Delhi, the latter being affected by the limited resolution of geostationary observations and rapid intensification constraints. In contrast, Cyclone Winston (2016) is \qty{43}{\hecto\pascal} weaker in MOTIS than in the RSMC Nadi Best Track, and $P_\mathrm{MOTIS}$ is much more aligned with other operational estimates, including ADT (\qty{914}{\hecto\pascal}), SATCON (\qty{925}{\hecto\pascal}), AMSU (\qty{913}{\hecto\pascal}), and SSMIS (\qty{936}{\hecto\pascal}). In these two cases, MOTIS appears to provide more accurate pressure estimates than the Best Tracks.

Taken together, these results suggest that MOTIS can substantially improve the quality and inter-basin consistency of Best Track $P_\mathrm{c}$ estimates for high-intensity TCs worldwide, with important implications for TC climatology, forecast verification, historical reanalysis, and risk modeling.

\section{Discussion}\label{sec:dis}

MODIS Thermal Infrared Sounding provides the most accurate $P_\mathrm{c}$ estimates for intense TCs compared to existing approaches, and the requirement of observing a clear eye is usually fulfilled for the strongest and most damaging TCs. The outstanding performance can be primarily attributed to the spatial resolution of MODIS, which is $\sim$30 times finer than that of microwave sounders and sufficient to resolve the compact cores of intense TCs. There are still some sources of error intrinsic to the theoretical model used, and we discuss these systematic errors in section~\ref{sec:dis}\ref{ssec:systematic}. Such an accurate intensity estimation tool will be valuable for reanalysis of past TCs, with implications discussed in section~\ref{sec:dis}\ref{ssec:role}. On the other hand, MODIS is nearing the end of its mission, leaving limited value for operational application. Therefore, in section~\ref{sec:dis}\ref{ssec:successors}, we discuss other infrared radiometers with the potential to provide accurate $P_\mathrm{c}$ estimates beyond the MODIS era.

\subsection{Systematic error}\label{ssec:systematic}

A major source of uncertainty in the current MOTIS algorithm comes from the estimation of warm-core top height, where the CDO temperature is currently used as a proxy. However, the correspondence between height and air temperature in TC environments varies from case to case. For a fixed CDO temperature, TCs with a higher local tropopause (often associated with larger TC size) can exhibit a higher geometric CDO height, because TC inner-core outflow can elevate the cold-point tropopause \citep{feng2021impacts, ditchek2017tropical}. Therefore, the current algorithm may underestimate the warm-core top height of larger TCs.

Indeed, we observe a trend of underestimating TCs with a large radius of outermost closed isobar (ROCI) measured in degrees latitude, by \qty{-0.4}{\hecto\pascal\per\degree}. Using a fixed 12--13$\degree$ latitude annulus $P_{\mathrm{env}}$ to replace POCI still leaves a trend of \qty{-0.3}{\hecto\pascal\per\degree}.

This trend could also partially explain the basin-specific bias mentioned in section~\ref{sec:results}\ref{ssec:model}, where estimates for WP storms are \qty{1.3}{\hecto\pascal} too weak and estimates for EP+CP storms are \qty{1.5}{\hecto\pascal} too strong in the observation sample, since the average ROCI of TCs in WP (8.5$\degree$) is significantly larger than the overall average (6.6$\degree$), whereas TCs in EP+CP are smaller (6.1$\degree$).

To address this systematic error, future improvements could incorporate a temperature-to-height conversion constructed from ERA5 data on complete model levels \citep{C3S_ERA5_complete_2023}.

In addition, studies have revealed a temporal lag between inner-core convective signals (e.g., CDO temperature) and warm-core development. As simulated by \citet{oyama2019relationship}, convective bursts typically precede peak warm-core temperature anomalies by 0--12~h. The Dvorak technique notes that convection in weakening storms can degenerate faster than the actual intensity \citep{velden2006dvorak}, which implies that the CDO-temperature proxy could make $P_\mathrm{MOTIS}$ too weak in weakening TCs. This is consistent with D-MINT/D-PRINT, which show that multi-hour IR histories can strongly influence CNN-based TC intensity estimates \citep{griffin2024predicting}. Although individual MODIS passes are too sparse to track CDO-temperature history, MOTIS could be improved by using lagged CDO-temperature information derived from geostationary IR imagery.

\subsection{Implications for TC-intensity estimation}\label{ssec:role}

Operationally, the maximum sustained wind ($V_{\max}$) tends to be the primary indicator of TC intensity, and $P_\mathrm{c}$ is often converted using wind-pressure relationships, such as the KZC relation \citep{knaff2007reexamination} applied in NHC Tropical Cyclone Reports \citep[e.g.,][]{cangialosi2024jova,bucci2026dora}. However, $V_{\max}$ suffers from definition inconsistencies and undersampling in measurements (see \citet{klotz2019sfmr}). $V_{\max}$ is variously defined over 1-, 3-, or 10-minute intervals for different RSMC/TCWCs, making it more difficult to compare across basins than $P_\mathrm{c}$. Direct wind measurements are also less accurate: \citet{landsea2013atlantic} quote a $\sim$11\,kt ($\sim$8\%) uncertainty in $V_{\max}$ even with aircraft observations, while $P_\mathrm{c}$ has a smaller estimated uncertainty of $\sim$\qty{4}{\hecto\pascal} ($\sim$5\%). The superior precision and universal definition of $P_\mathrm{c}$ mean that remote sensing tools have a higher precision ceiling for estimating pressure than for estimating wind, and our MOTIS framework provides a key step toward this ceiling.

Currently, $P_\mathrm{best}$ from RSMC/TCWCs still suffers from uncertainties caused by inconsistencies in the analysis tools and wind-pressure relationships used. Section~\ref{sec:results}\ref{ssec:reanalysis} discusses some major discrepancies between Best Track and MOTIS and demonstrates that they likely arise from limitations in Best Track quality. Inconsistencies in Best Track quality mainly arise from the regional dependence of analysis tools (e.g., the Dvorak/ADT workflow depends on the quality of geostationary observations \citep{velden2006dvorak,olander2019advanced}) and agency-specific conventions. MOTIS could help improve the coherence and quality of Best Track estimates, since it offers global coverage and high accuracy up to a zenith angle of $64\degree$ (see Figure \ref{fig:error distribution}).

\subsection{Future successor: Geostationary sounders}\label{ssec:successors}

The MODIS mission is planned to end in 2027. To the best of our knowledge, however, no operational infrared imagers have sufficient $\mathrm{CO_2}$ sounding capability. For example, as the nominal successor to MODIS, VIIRS has no $\mathrm{CO_2}$ absorption bands; geostationary imagers such as ABI and AHI have only one $\mathrm{CO_2}$ absorption band at \qty{13.3}{\micro\meter}, which mainly responds to the lower troposphere but fails to sample the bulk TC warm core.

New-generation geostationary hyperspectral IR sounders have greatly improved spatial resolution and are promising successors to MODIS for providing $P_\mathrm{c}$ estimates. Unlike polar-orbiting IR sounders with limited IFOV and spatial sampling due to integration time constraints, geostationary IR sounders can achieve much higher spatial resolution and observation frequency. For example, the IRS on MTG-S1 and the GIIRS on FY-4C (both satellites launched in 2025) achieve a resolution of \qty{4}{\kilo\meter} at nadir, comparable to the imager resolution of satellites such as the MTSAT series (2005--2015) and GOES-N series (2010--2018). Future instruments, including GHMS on Himawari-10 (planned launch in 2029) and GXS on GeoXO-1 (planned launch in 2032), will provide more extensive coverage across TC-prone basins.

Geostationary IR sounders still have coarser spatial resolution than MODIS, but they offer several advantages that are valuable to $P_\mathrm{c}$ estimation. First, these hyperspectral sounders have thousands of spectral bins, which enable more accurate temperature profile retrieval in the eye. Second, the spectral range of IR sounders extends beyond Band~36's \qty{14.2}{\micro\meter} into a wavelength region with even stronger $\mathrm{CO_2}$ absorption, sensing the warm core near the top of the troposphere ($\sim$\qty{100}{\hecto\pascal}) and fundamentally resolving the issue of warm-core top height estimation. Third, while polar-orbiting satellites revisit a given location in the tropics at most twice a day, geostationary satellites have continuous coverage and can provide temporally dense data that facilitate both model fitting and operational usage. Fourth, for satellites that carry both a sounder and an imager (e.g., Himawari-10), if the sounder cannot fully resolve the warmest region in the eye, the higher-resolution imager can be co-registered to perform super-resolution or to provide undersampling correction. Hence, algorithms based on these future geostationary IR sounders could provide TC intensity estimates comparable to or even better than those achieved by MOTIS.

\section{Conclusions}

We present MODIS Thermal Infrared Sounding (MOTIS), a novel technique for estimating TC central pressure from high-resolution MODIS observations of the warm-core structure in mature TC eyes. For high-intensity TCs with clear eyes, MOTIS achieves an accuracy that outperforms all existing tools and could reduce Best Track uncertainty by $\sim$50\% in cases without direct observations. We have also created a large dataset of MOTIS estimates of TC central pressure, spanning 2002--2025 and including more than 1000 high-accuracy pressure estimates for TCs with clear eyes (\texttt{https://motis.dapiya.cn}). MOTIS estimates can support global TC studies and be assimilated into numerical models to improve retrospective simulations of past storms. In addition, MOTIS estimates can shed light on differences in RSMC/TCWC standards and have the potential to improve the TC dataset for climatological studies.

Although MODIS is planned to be discontinued by 2027, a new generation of geostationary IR hyperspectral sounders will soon be available for observing TCs. They have strong potential for developing algorithms similar to MOTIS for TC intensity estimation, and it is important that this potential is not overlooked, as MODIS's potential was for the past 20 years. The advantages of the MOTIS framework we presented lie in MODIS's spectral properties and spatial resolution, not in the analysis technique; therefore, developing advanced methods that fully ingest two-dimensional data from MODIS or future sounders, similar to D-MINT \citep{griffin2024predicting}, could provide even better remote sensing estimates of TC intensity in the absence of \textit{in situ} observations.

\clearpage

\acknowledgments
The authors thank the data providers and operational satellite-intensity teams whose products were used in this study. This work was supported in part by NOAA Research under award NA25NESX432C0001, ``CIMSS Research Towards the SATellite CONsensus (SATCON) Tropical Cyclone Intensity Estimation Algorithm for Geo/Leo Weather Satellite Data Usage FY25.'' The authors declare no conflicts of interest.

\datastatement
The input datasets used in this study are publicly available from the sources cited in the text. The derived MOTIS estimate dataset is available at \texttt{https://motis.dapiya.cn}. Per-pass diagnostic plots are being prepared and will be added to the same website.

\appendix[A]
\appendixtitle{Selected Intense Tropical Cyclones}

\begin{table}[h]
\caption{MOTIS estimates for selected intense TCs$^{\rm a}$.}
\begin{center}
\scriptsize

\begin{tabular}{lm{1.75cm}<{\centering}m{0.4cm}<{\centering}m{0.5cm}<{\centering}m{0.6cm}<{\centering}m{0.5cm}<{\centering}m{0.5cm}<{\centering}}
\topline
Name & Time (UTC) & Basin & $\Delta T_\mathrm{B36}$ (\unit{\kelvin}) & $T_{\mathrm{CDO}}$ (\unit{\degreeCelsius}) & $\Delta P$ (\unit{\hecto\pascal}) & $P_\mathrm{c}$ (\unit{\hecto\pascal}) \\
\midline
Meranti  & 2016.09.13 17:15 & WP & 11.1 & -75.5 & 137.5 & 870.1 \\
Meranti  & 2016.09.13 14:30 & WP & 11.7 & -76.0 & 136.9 & 871.9 \\
Nepartak & 2016.07.06 04:50 & WP &  8.7 & -78.2 & 131.6 & 875.0 \\
Megi     & 2010.10.17 17:05 & WP & 10.8 & -77.7 & 131.0 & 877.0 \\
Haiyan   & 2013.11.11 16:40 & WP &  8.6 & -84.5 & 120.7 & 885.5 \\
Megi     & 2010.10.17 14:20 & WP & 11.1 & -77.3 & 123.0 & 885.8 \\
Bolaven  & 2023.10.11 16:00 & WP &  9.0 & -76.4 & 123.5 & 886.5 \\
Patricia & 2015.10.23 17:30 & EP &  9.3 & -77.3 & 124.3 & 886.6 \\
Olaf     & 2005.02.16 01:05 & SP &  7.1 & -81.8 & 113.8 & 887.7 \\
Meranti  & 2016.09.12 13:50 & WP &  9.1 & -77.0 & 122.7 & 887.9 \\
Melissa  & 2025.10.28 14:45 & AL &  9.6 & -78.6 & 115.5 & 894.9 \\
Gafilo   & 2004.03.06 06:55 & SI & 10.1 & -77.2 & 113.1 & 897.3 \\
Phailin  & 2013.10.10 19:30 & IO &  7.8 & -77.1 & 110.8 & 899.9 \\
Monica   & 2006.04.23 13:10 & AU &  9.3 & -76.8 & 107.3 & 902.2 \\
\botline
\end{tabular}
\end{center}
\footnotesize{$^{\rm a}$ Included cases either have $\Delta P > 120$\,hPa or represent the largest MOTIS-estimated $\Delta P$ in their respective basins. Only good-eye samples are included.}
\label{tab:highlighted-tc-cases}
\end{table}

Table~\ref{tab:highlighted-tc-cases} summarizes MOTIS estimates for selected intense TC cases. Note that the satellite passes do not necessarily coincide with the lifetime peak intensity of each TC. A complete dataset of all available MOTIS estimates is provided at \texttt{https://motis.dapiya.cn}.

\appendix[B]
\appendixtitle{Band Specification}

\begin{table}[h]
\begin{center}
\footnotesize
\caption{MODIS IR bands used in this study (resolution = 1\,km).}
\label{tab:bands}
\begin{tabular}{m{0.5cm}<{\centering}m{1.0cm}<{\centering}m{2.2cm}<{\centering}m{2.5cm}<{\centering}}
\toprule[1pt]
Band & $\lambda_0$ (\si{\micro\meter}) & Band type & Role in this study \\
\midrule
24 & 4.466  & $\mathrm{CO_2}$ absorption & Warm-core sounding \\
27 & 6.715  & H$_2$O absorption   & Absorption context \\
31 & 11.030 & IR window               & Convection reference \\
32 & 12.020 & IR window               & Crosstalk reference \\
33 & 13.335 & \multirow{4}{*}{$\mathrm{CO_2}$ absorption} & \multirow{4}{*}{Warm-core sounding} \\
34 & 13.635 &  &  \\
35 & 13.935 &  &  \\
36 & 14.235 &  &  \\
\bottomrule[1pt]
\end{tabular}
\end{center}
\end{table}

Table \ref{tab:bands} summarizes the eight MODIS IR bands used in this study.
Bands~24--25 and Bands~33--36 respond to $\mathrm{CO_2}$ absorption.

Band~25 has the highest surface transmittance among the six $\mathrm{CO_2}$ bands, and is not adopted in MOTIS due to its sensitivity to low-level clouds and sunlight.

Three additional TIR bands are adopted in this study: Band~27, which helps estimate $\mathrm{H_2O}$ absorption in Bands~33--36; Band~31, which is used to measure the central dense overcast (CDO) temperature of TCs; and Band~32, which is used both as the reference band for crosstalk removal (appendix~C, section~\ref{sssec:crosstalk}) and as a split-window band against Band~31 to estimate cirrus cloud absorption in Band~24 and Bands~33--36.

\appendix[C]
\appendixtitle{MODIS preprocessing}

We detail the following preprocessing steps: (a) crosstalk removal, (b) mirror side correction, (c) zenith angle adjustment, and (d) noise subtraction.

\subsection{Crosstalk removal}\label{sssec:crosstalk}

The $\mathrm{CO_2}$ absorption bands are affected by crosstalk, in which signals from other bands contaminate individual detectors and produce stripe patterns. Because some artifacts remain in MODIS C6.1, we apply additional band-specific corrections.

\begin{figure}[htbp]
    \centering
	\noindent\includegraphics[width=\columnwidth]{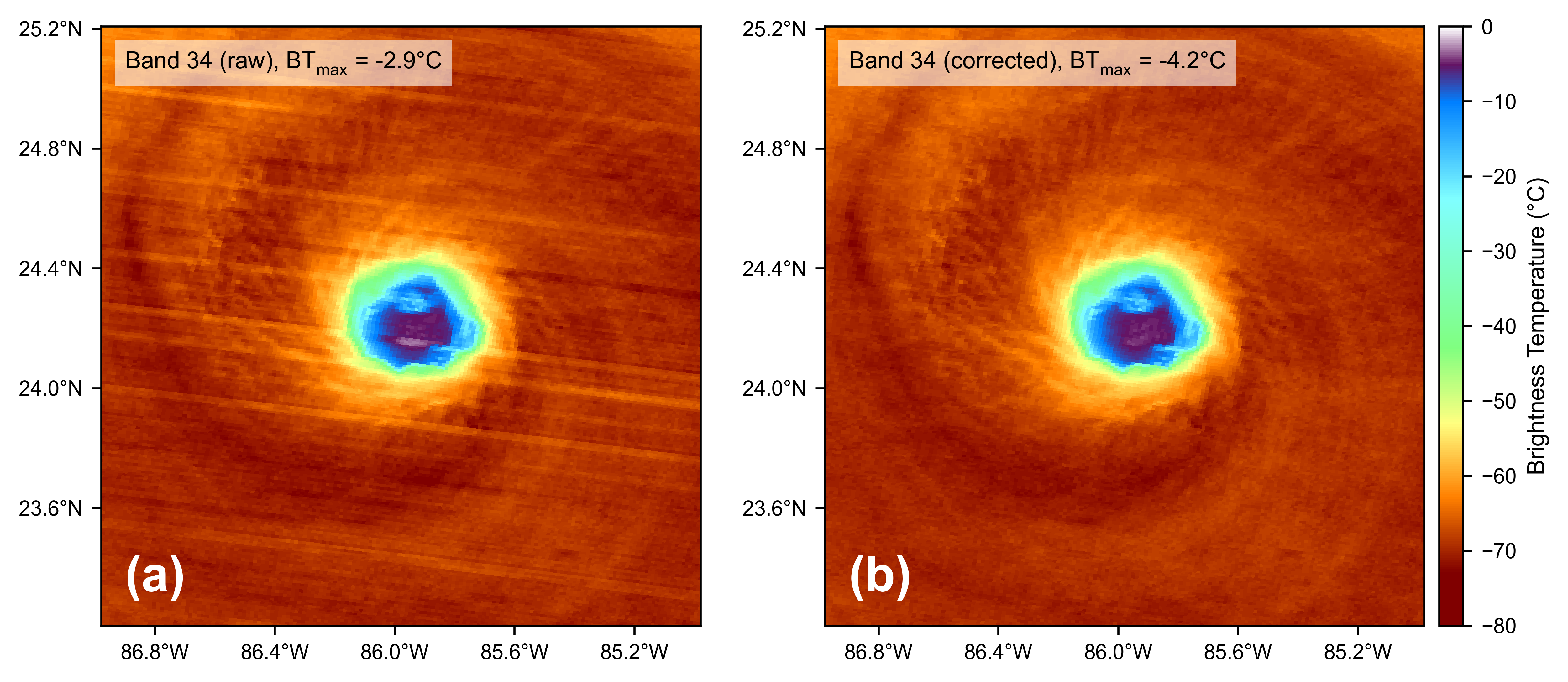}
	\caption{Zoomed-in Terra Band~34 imagery of Hurricane Rita (2005) at UTC 16:10 on Sep 21. The left and right panels show the original and crosstalk-corrected $BT$, respectively.}\label{figure:crosstalk}
\end{figure}

Band~24 is susceptible to electronic crosstalk leaking from Band~26 \citep{keller2017aqua}, which causes detector 1 to have significantly higher $BT$ or even NaN values during daytime. To address this issue, we predict Band~24 digital number (DN) values using polynomial regression on a reference band (Band~31) with negligible crosstalk and subtract the predicted values from the actual Band~24 values. We perform nearest-neighbor filling for detector 1 on the subtracted image and then add the predicted values back to reconstruct destriped Band~24. This procedure helps keep the cloud structure intact while removing crosstalk.

For Bands~33--36, Terra MODIS is affected by optical leakage from Band~31, which can imprint displaced scene structure on multiple detectors. Although C6.1 includes this correction \citep{xiong2005modis}, residual contamination remains. We apply a similar reference-band regression using Band~32 and different treatments to the subtracted image. For Bands~35--36, we apply a median filter with a window of $3\times1$ pixels along the cross-detector direction. For Bands~33--34, where significant multidetector crosstalk remains, we apply a low-pass filter that subtracts a $1\times31$ row-smoothed residual after removing scene structure with a $19\times1$ median background. Figure \ref{figure:crosstalk} shows an example of Band~34 crosstalk removal.

\subsection{Mirror side correction}\label{sssec:mirror}

MODIS C6.1 has known issues with mirror side differences, as detailed in \citet{angal2023status}. We examined the mirror side difference of Bands~33--36 (photoconductive) and Band~24 (photovoltaic) across the operational period of both satellites, using cloud-masked ocean scenes within $\pm 20\degree$ latitude, which matches typical TC environments.

We found an increasing trend of Band~36 mirror side difference on Aqua MODIS (growing linearly from \qty{0.024}{\watt\per\square\meter\per\micro\meter\per\steradian} in 2002 to \qty{0.038}{\watt\per\square\meter\per\micro\meter\per\steradian} in 2025 for a zenith angle of $50^\circ$), whereas the mirror side difference is insignificant on Terra MODIS. Similar but more tentative trends were found for Bands~33--35 as well; Band~24 did not show significant mirror side differences on either satellite. Therefore, we adopted the mirror side correction obtained on Band~36 and applied it across Bands~33--36.

\subsection{Zenith angle adjustment}\label{ssec:zenith}
Across the MODIS swath, the instrument's zenith angle can be as large as $65\degree$, and the raw $BT$ is typically significantly colder at large zenith angles. We perform zenith angle adjustment for both $\Delta BT$ retrieval and CDO temperature estimation.

\begin{figure}[htbp]
    \centering
    \noindent\includegraphics[width=0.8\columnwidth]{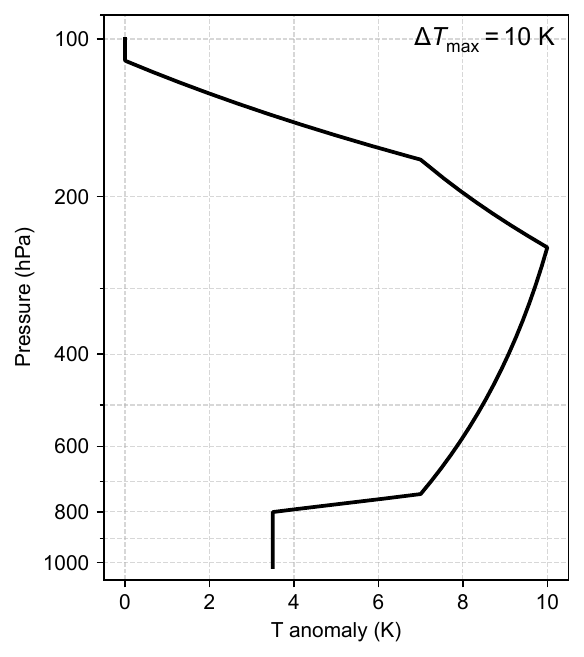}
	\caption{The vertical warm-core anomaly profile assumed in zenith angle adjustment, adapted from \protect\citet{stern2012height}.}\label{figure:profile}
\end{figure}

The cooling of $BT$ is not fixed for a given angle but depends on the actual atmospheric profile. Some studies have used latitude and seasonality as proxies \citep{elmer2016limb}, but such an approach is not robust inside TC eyes, since the atmospheric profile is heavily modified compared to the ambient environment. Therefore, we assume a TC warm-core anomaly proportional to the profile shown in Fig.~\ref{figure:profile}, which is adapted from \citet{stern2012height}. The magnitude of anomaly is iteratively retrieved with RTTOV by scaling the profile according to $\Delta BT$, simulating the zenith adjustment, and updating $\Delta BT$ until convergence.

For CDO temperature, RTTOV simulations are no longer reliable without detailed cloud models. We therefore empirically fit the zenith angle adjustment using CDO temperature from the ADT archive \citep{olander2019advanced} as a reference, yielding the following $BT$ adjustments:

\begin{equation}
    \begin{split}
        \Delta T_\mathrm{CDO\ adj.}(\theta) = (34.2+0.3579\times T_\mathrm{CDO})\tan^2{\frac{\theta}{2}} \\ + \theta \times \begin{cases} 0.00458 ~ \mathrm{K \degree^{-1}} & \mbox{if Aqua} \\ 0.00794 ~ \mathrm{K \degree^{-1}} & \mbox{if Terra} \end{cases}
    \end{split}
\end{equation}
where the linear $\theta$ term accounts for additional mirror side effects.

Together with mirror side corrections, zenith angle adjustments keep $\Delta BT$ and $T_\mathrm{CDO}$ insensitive to satellite viewing angles, making MOTIS estimates robust even at large zenith angles, as shown in Figure \ref{fig:error distribution} (c).

\subsection{Noise}
Band~36 has especially low radiance and $BT$, and is subject to greater random noise than other bands, which can inflate the maximum observed $BT$ (${T}_{\mathrm{max}}$). To estimate the noise effect, ${T}_{\mathrm{max}}$ in Band~36 is compared with the $90^\mathrm{th}$ percentile of $BT$ $(T_\mathrm{P90})$ for pixels having $T>{T}_{\mathrm{max}}-2$. Band~33 is used as the reference band, as it is the $\mathrm{CO_2}$ band with the highest radiance and presumably the least random noise.

The Band~36 noise is then defined as:
\begin{equation}
    T_\mathrm{noise}=(T_\mathrm{max, B36}-T_\mathrm{P90, B36})-(T_\mathrm{max, B33}-T_\mathrm{P90, B33})
\end{equation}

We then subtract $0.5T_\mathrm{noise}$ from the Band~36 $BT$ anomaly ($\Delta T_\mathrm{B36}$). We do not subtract the full $T_\mathrm{noise}$ because part of the calculated $T_\mathrm{noise}$ could reflect genuine variation in the temperature distribution rather than random noise.

\appendix[D]
\appendixtitle{Additional regression parameters}

To address factors that affect the measured $\Delta BT$, additional regression parameters are used in the multiple linear regression, as described below.

\subsection{Brightness Temperature Anomaly Difference}\label{ssec:btd}

The vertical distribution of the TC warm core is captured using the ${\Delta T}$ difference between neighboring bands, a design that reduces collinearity in the regression compared to using each band's $BT$ anomaly directly.

We adopt the difference between ${\Delta T}_\mathrm{B36}$ and ${\Delta T}_\mathrm{B35}$ and that between ${\Delta T}_\mathrm{B35}$ and ${\Delta T}_\mathrm{B34}$, which are representative of the vertical gradient of the warm core in the upper and mid-level troposphere. Specifically, the parameters are defined as follows:

\begin{equation}
T_\mathrm{dif}=\frac{\Delta BT - \Delta BT'}{\mathrm{max}(\Delta BT+ \Delta BT',9) / 2}
\end{equation}

where the $\Delta BT$ difference is normalized by the average of two bands, guarded by a minimum temperature to prevent extreme values in case of cloud obscuration.

\subsection{Pixel count}\label{ssec:pixel}

The observed eye temperature is reduced when the true eye temperature is undersampled. For MODIS, this resolution effect is estimated to be inversely proportional to the effective eye size $r_\mathrm{eff}$, down to a certain minimum size. $r_\mathrm{eff}$ was calculated as follows:
\begin{equation}
    r_\mathrm{eff}=\sqrt{\mathrm{max}(n_\mathrm{pix},n_\mathrm{min})}
\end{equation}

where $n_\mathrm{pix}$ is defined as the number of Band~35 eye pixels within \qty{1}{\kelvin} of $T_\mathrm{max,B35}$. Band~35 is adopted because it is the highest-altitude band without significant noise. The \qty{1}{\kelvin} threshold was chosen because it empirically minimized the error in $\Delta P_\mathrm{MOTIS}$. Extremely small $r_\mathrm{eff}$ is avoided by setting $n_\mathrm{min}=8$, which empirically produced the best $\Delta P_\mathrm{MOTIS}$.

\subsection{$\mathrm{H_2O}$ absorption band}\label{ssec:h2o}

Besides the $\mathrm{CO_2}$ absorption bands, we also included the $BT$ anomaly of the $\mathrm{H_2O}$ absorption band (Band~27), which reflects both the temperature and $\mathrm{H_2O}$ concentration in the middle and upper troposphere.

Bands~33--36 are primarily sensitive to $\mathrm{CO_2}$ absorption, with some contribution from $\mathrm{H_2O}$. Band~27 is dominated by $\mathrm{H_2O}$ absorption, whereas Band~24 is sensitive to both $\mathrm{CO_2}$ and $\mathrm{N_2O}$. Unlike $\mathrm{CO_2}$ and $\mathrm{N_2O}$, atmospheric $\mathrm{H_2O}$ decreases rapidly with height, giving Band~27 a unique narrow-peaked weighting function that helps constrain the vertical warm-core profile. Meanwhile, $\mathrm{H_2O}$ absorption may introduce unwanted $BT$ signals to Bands~33--36 caused by humidity differences. MOTIS addresses any potential influence from $\mathrm{H_2O}$ variability by incorporating Band~27 into the regression.

\subsection{Cirrus cloud absorption}\label{ssec:cirrus}

Cirrus absorption can reduce $BT$ in obscured eyes. To estimate the absorption, we adopt the difference between Band~31 and Band~32, which has long been used in cirrus identification (see \citet{strabala1994cloud}) because of the wavelength dependence of ice particle absorption \citep{warren2008optical}.

\begin{figure}[htbp]
    \centering
	\noindent\includegraphics[width=\columnwidth]{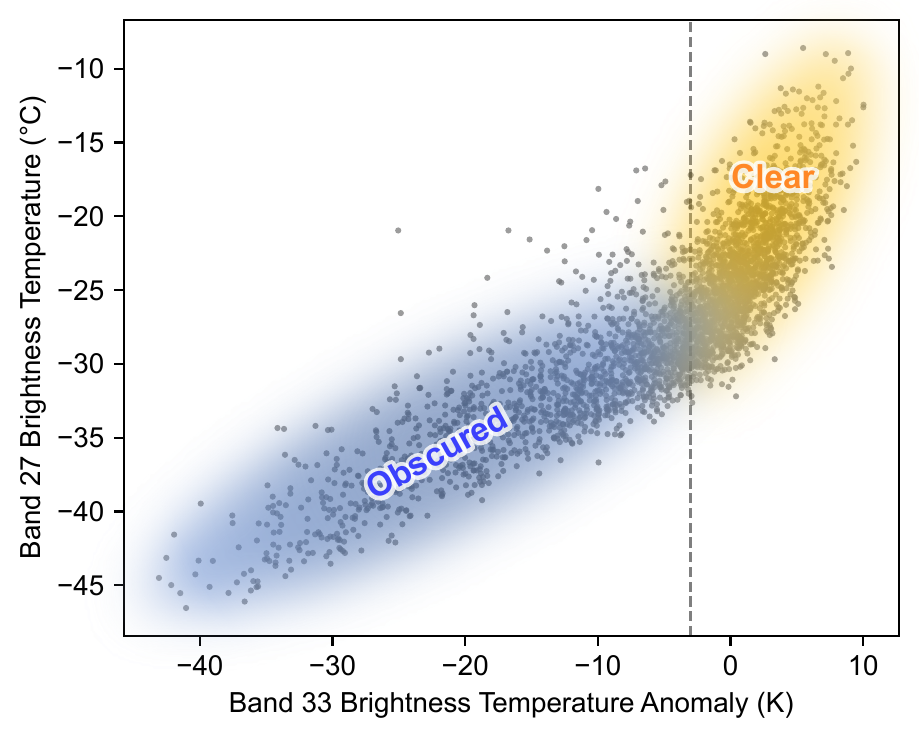}
	\caption{Scatterplot of $T_\mathrm{B27}$ against $\Delta T_\mathrm{B33}$. The vertical dotted line shows $\Delta T_\mathrm{B33} = -3~\mathrm{K}$, which splits the samples into two regions with different slopes. Manual inspection suggests that the two regions correspond to generally clear eyes and eyes with varying levels of obscuration.}
    \label{Tbd33 vs Tb27}
\end{figure}

In addition, Figure \ref{Tbd33 vs Tb27} shows a knee point at ${\Delta T}_\mathrm{B33}=$\qty{-3}{\kelvin}, likely caused by cirrus absorption. This suggests that ${\Delta T}_\mathrm{B33}$ below this threshold is indicative of cirrus clouds. Therefore, we use ${\Delta T}_\mathrm{B33, trunc.} = \mathrm{max}(-3-{\Delta T}_\mathrm{B33},0)$ as another parameter that compensates for cirrus absorption.

\subsection{Instrument bias}\label{ssec:bias}

We identify two sources of instrument bias: the Band~24 daytime reflection contribution of Aqua MODIS and the overall spectral response bias of Terra MODIS relative to Aqua.

Band~24 has a wavelength centered at \qty{4.5}{\micro\meter}, where reflected sunlight can contribute to $BT$. While Terra shows a uniform all-day $\Delta T_\mathrm{B24}$ distribution, Aqua $\Delta T_\mathrm{B24}$ is generally higher during daytime. The bias is particularly significant for passes with high zenith angles and low Band~31 temperature, corresponding to scenarios with significant reflection from the eyewall and in-eye clouds. To address this issue, we first removed the uniform daytime bias of Aqua (daytime: \qty{0.5}{\kelvin}; nighttime \qty{-0.5}{\kelvin}). We then include a parameter $T_{\mathrm{rfl.}} = (T_\mathrm{base}-T_\mathrm{B31})\tan^2{\frac{\theta}{2}}$ in the regression to approximate cloud reflection, since we lack information to physically model the clouds.

In addition, the spectral responses of MODIS on Aqua and Terra are close but not identical. Therefore, a parameter ${\Delta T}_\mathrm{B34}$ is included for Terra MODIS, accounting for its difference from Aqua. Band~34 is chosen because its residual crosstalk on Terra, a significant source of the instrument bias, is the most severe.

\bibliographystyle{ametsocV6}
\bibliography{references}

\end{document}